# A dynamical stability limit for the charge density wave in $K_{0.3}MoO_3$

R. Mankowsky[1,2], B. Liu[1], S. Rajasekaran[1], H. Liu[1], D. Mou[3], X. J. Zhou[3], R. Merlin[4], M. Först[1] and A. Cavalleri[1,2,5]

[1]Max Planck Institute for the Structure and Dynamics of Matter, Hamburg, Germany
[2]University of Hamburg, Hamburg, Germany
[3]Beijing National Laboratory for Condensed Matter Physics, Institute of Physics, Chinese Academy of Sciences, Beijing, China
[4]Department of Physics, University of Michigan, Ann Arbor, Michigan, USA
[5]Department of Physics, Oxford University, Clarendon Laboratory, Oxford, UK

We study the response of the one-dimensional charge density wave in $K_{0.3}MoO_3$ to different types of excitation with femtosecond optical pulses. We compare the response to direct excitation of the lattice at mid-infrared frequencies with that to the injection of quasi-particles across the low-energy charge density wave gap and to charge transfer excitations in the near infrared. For all three cases, we observe a fluence threshold above which the amplitude-mode oscillation frequency is softened and the mode becomes increasingly damped. We show that all the data can be collapsed onto a universal curve in which the melting of the charge density wave occurs abruptly at a critical lattice excursion. These data highlight the existence of a universal stability limit for a charge density wave, reminiscent of the empirical Lindemann criterion for the stability of a crystal lattice.

One-dimensional charge-density waves (CDWs) are a prototypical example of a broken symmetry state, in which the energy is lowered by the modulation of the conduction electron density at a wave vector $q_{CDW}$ that nests two regions of the Fermi surface [1]. This modulation in the real space charge distribution is typically also associated with a distortion of the lattice [2] and with the appearance of new excitations that can be observed in optical or Raman spectroscopy. Charge Density Waves are especially interesting in systems in which nesting occurs between two different bands, for which the order is often incommensurate with the lattice, resulting in highly anomalous and nonlinear DC conductivity.

Here, we study the response of the one-dimensional charge density wave material $K_{0.3}MoO_3$ (blue bronze) to short pulse optical excitation, which so far has been investigated extensively at near-infrared wavelengths [3,4,5]. In these studies, a collapse of the CDW gap [6] and the rearrangement of the lattice along the coordinate of the amplitude mode [7] were reported.

We seek to provide a new perspective into these dynamics by inducing melting of the same charge density wave by three different types of optical stimulation. We study excitation of the lattice alone, of charge quasi-particles immediately above the low energy CDW gap and of charges across a high-energy charge transfer resonance. The third mechanism is the one studied in previous experiments and is analyzed here as a reference point. We find evidence for a universal dynamical instability that occurs always at specific lattice displacement, which we estimate to be of approximately 20% of the CDW-induced equilibrium lattice distortion.

$K_{0.3}MoO_3$ is a prototypical one-dimensional charge-density wave material, made up of conducting chains of corner-sharing $MoO_6$ octahedra along the $b$ direction of its monoclinic crystal structure (see Fig. 1a). Below $T_C$ = 183 K, incommensurate CDW order with an energy gap $2\Delta$ = 130 meV develops at a temperature dependent wave vector $q_{CDW} = \left(1, q_b, \overline{0.5}\right)$, with $q_b$ = 0.748 at 100 K [8]. The corresponding lattice distortion is mostly transverse, involving Mo-O bond length changes perpendicular to the *b* direction on the order of 5 pm [9,10]. The Mo-O chains are separated from each other by sheets of potassium atoms making this material insulating in the perpendicular directions with an energy gap of 1.3 eV.

Due to their strong coupling to the electronic order, several phonon modes are folded back to the Brillouin zone center and appear in infrared [11] and Raman spectra [12,13]. The amplitudes, frequencies and damping of these collective excitations hold important information on the density of the ordered charges [14].

Our experiments were performed on a $K_{0.3}MoO_3$ single crystal, cooled to 20 K base temperature, well below the transition temperature. Mid-infrared pulses of 150 fs duration, with few-µJ of energy and tunable between 6 and 15 µm wavelength, were used to resonantly excite the infrared-active 1000 cm$^{-1}$ Mo-O stretching mode polarized along the [-201] crystallographic direction (see Figures 1b and 2a). Significant coupling of this phonon mode to the CDW order is known to occur also at equilibrium, as evidenced by a prominent reshaping in the optical conductivity across the metal-to-CDW transition [15].

In a parallel set of experiments, the charge density wave itself was optically excited in the same wavelength range, using mid-infrared pulses that were polarized along the [010] axis and tuned to the CDW gap (red area in Fig.2b). For comparison, melting of the condensate was also studied using near-infrared pulses tuned to wavelengths above the 1.3-eV charge transfer gap and polarized along [-201], perpendicular to the $MoO_6$ octahedra chains (grey area in Fig. 2a).

In all three cases, the $K_{0.3}MoO_3$ reflectivity at 800 nm wavelength was measured as a function of time delay using 35-fs pulses, polarized along the chain direction [010]. As the penetration depths of the different pump pulses were in all cases larger than the one of the 800-nm probe pulses, the dynamical response in the sample reflectivity can directly be compared.

Figure 2c shows the time-resolved reflectivity changes for the three different excitation conditions. We report here excitation at low fluence, far below the charge density wave melting threshold [6]. In all three cases, we found a prompt reduction and double-exponential recovery in the reflectivity ascribed to electronic dynamics in earlier studies [3,4,6]. Coherent oscillations of multiple zone-folded Raman modes were observed along with the incoherent response, dominated by the charge density wave amplitude mode at 1.68 THz. By subtracting a double-exponential fit to the incoherent electronic background, we obtained the purely oscillatory response (Figure 3a), which we fitted by the sum of three exponentially decaying oscillations with frequencies of the amplitude mode at 1.68 THz and the first doublet of zone-folded Raman modes at 2.23 and 2.56 THz (Figure 3b).

The absolute phase of these oscillations was derived by extrapolating the fit to the pump pulse arrival time at $t = 0$. To precisely determine the zero time delay in the different experimental conditions, we simultaneously measured a four-wave mixing signal at frequency $2\omega_{probe} + \omega_{pump}$, which directly yielded the cross correlation between the pump and probe pulses (see Figure 2c).

The direct excitation of the Mo-O lattice mode resulted in coherent excitation of the amplitude mode with a sine phase, indicative of impulsive excitation, for which the driving force acts on a timescale short compared to the period of the amplitude mode of 600 fs. Assuming lattice anharmonicities as driving force, such impulsive excitation can be understood if the lifetime of the resonantly excited infrared-active mode is equal to or shorter than the period of the anharmonically coupled amplitude mode [16]. We refer here to previous work on the nonlinear coupling between infrared and Raman modes by nonlinear phononics. In agreement with this picture, the amplitude of the coherent oscillations was found to peak when tuning the frequency of the pump pulses to the resonance frequency of the infrared-active mode as shown in Fig. 3c [17].

Cosine phase oscillations were measured for electronic excitations across the [-201] charge transfer (black curve) and the [010] CDW gap (red curve), suggestive of displacive excitation of the coherent amplitude mode [18,19]. For these two types of excitation the amplitude of the coherent oscillations closely follows the optical conductivity as function of pump pulse frequency (Fig. 3d,e).

Figure 4 displays the fluence-dependence of amplitude, frequency and scattering rate of the CDW amplitude mode coherent dynamics. All three types of excitation caused the same nonlinear behavior in the optical response (Fig. 4a,b,c), showing first a linear increase of the oscillation amplitude, followed by a saturation concomitant with a frequency softening and a nonlinear increase in scattering rate. The threshold value of the excitation energy density, calculated by taking into account reflectivity losses and the penetration depth of the pump pulses, is approximately 3J/cm$^3$ for phonon excitation and 30J/cm$^3$ for both electronic excitations (see upper panels of Figure 4). The latter two densities are similar to values previously obtained in 800nm pump experiments (Ref. 6).

Note also that despite the different threshold values for the three excitation conditions, saturation of the oscillation amplitude was always observed at a change in reflectivity $\Delta R/R_0$ ~2%, which was probed by the same 800 nm wavelength pulses. As the change in optical reflectivity along a coherent phonon coordinate is linearly proportional to the real space atomic displacement, the observation reported here indicates that in all three cases CDW melting occurs for the same displacement of the amplitude mode [19]. This amplitude can be quantitatively estimated by comparing previously measured changes in 800-nm optical reflectivity (from Ref. 6) with femtosecond x-ray diffraction measurements (from Ref. 7) induced by the same excitation. According to this estimate, the 2% $\Delta R/R_0$ at the CDW melting threshold corresponds to a coherent displacement of the amplitude mode of 22% of the static lattice distortions that accompany the equilibrium CDW-to-metal transition.

The frequency and relaxation was then plotted as a function of the coherent oscillation amplitude. As shown in Figure 5, both mode softening and scattering rate fall on a single curve, which indicates abrupt dynamical melting of the charge density wave phase at a critical lattice displacement of approximately 20% of the equilibrium displacements.

These data provide a new view for the dynamical melting of electronic order, identifying a stability limit for one-dimensional charge density waves. This effect is more reminiscent of the Lindemann criterion for structural melting than of a second order transition, which proceeds by mode softening and divergence of fluctuations. This appears to be also in agreement with studies of charge density wave melting in two dimensional compounds like $TaS_2$, where the collapse of the bandgap measured by photoemission [20] and optical spectroscopy [21] appears to be accompanied by melting of the charge density wave [22] at an incomplete relaxation of the lattice [23].

Note also that similar critical displacements could be extracted from previous measurements of the solid-to-liquid transition in Bismuth [24], which included mode softening [25] above a critical photo-excitation threshold. The present data restricts the observation to a one-dimensional system and an additional benchmark for theoretical work, toward a general theory for non-equilibrium phase transitions.

The observations reported here provide also interesting perspective for experiments, in which melting of competing charge density waves with light enhance another order. Indeed, in the case of high-temperature cuprate superconductors, enhancement of superconducting coupling [26, 27, 28, 29] was found to follow optical excitation, accompanied by melting of charge order, which appears to take place both for excitation at visible frequency [30] as well as with mid infrared pulses [31, 32].

**FIGURES**

**Figure 1**

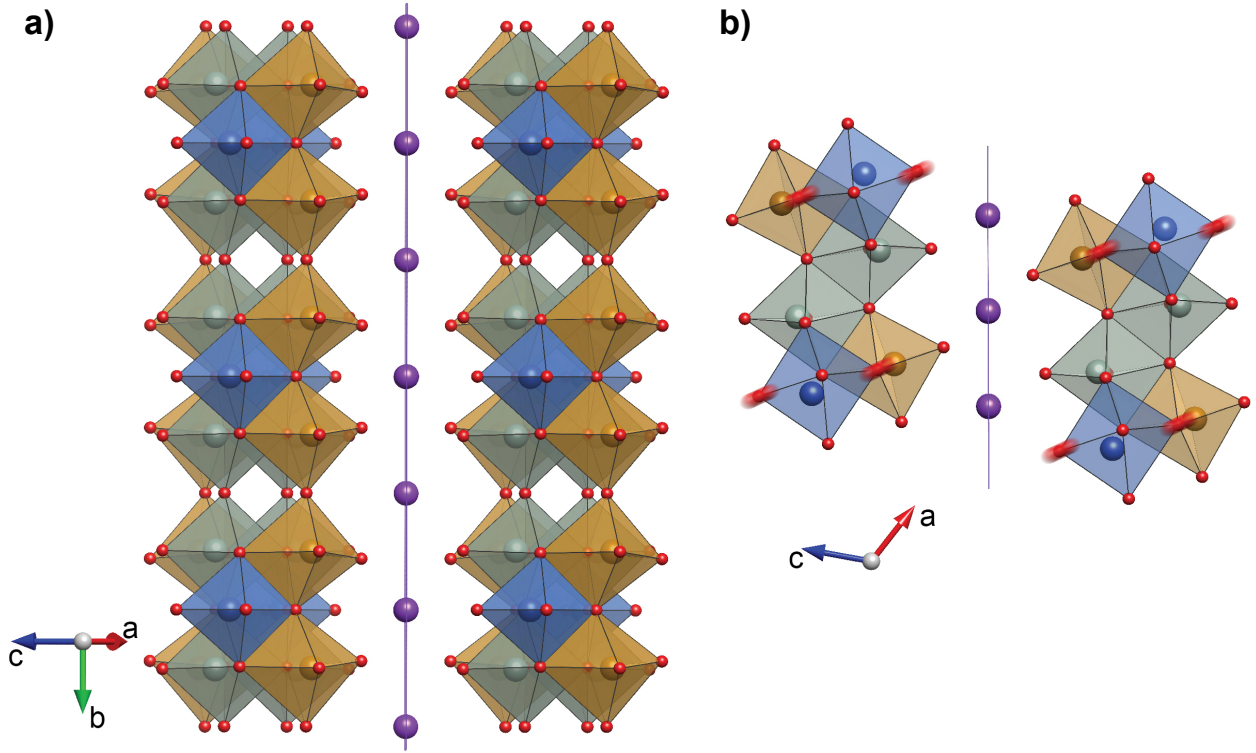

**Fig. 1:** (a) Monoclinic crystal structure of $K_{0.3}MoO_3$. Corner-sharing $MoO_6$ octahedra form chains along the [010] direction that are separated by sheets of potassium atoms (purple). Below $T_C$ = 183 K, incommensurate charge density wave order develops at a wave vector $q_{CDW} = (1, q_b, \overline{0.5})$, with $q_b$ = 0.748 at 100 K. (b) Top view of the chains. The resonantly excited phonon mode comprises Mo-O stretching along the insulating [-201] direction, perpendicular to the chains.

**Figure 2**

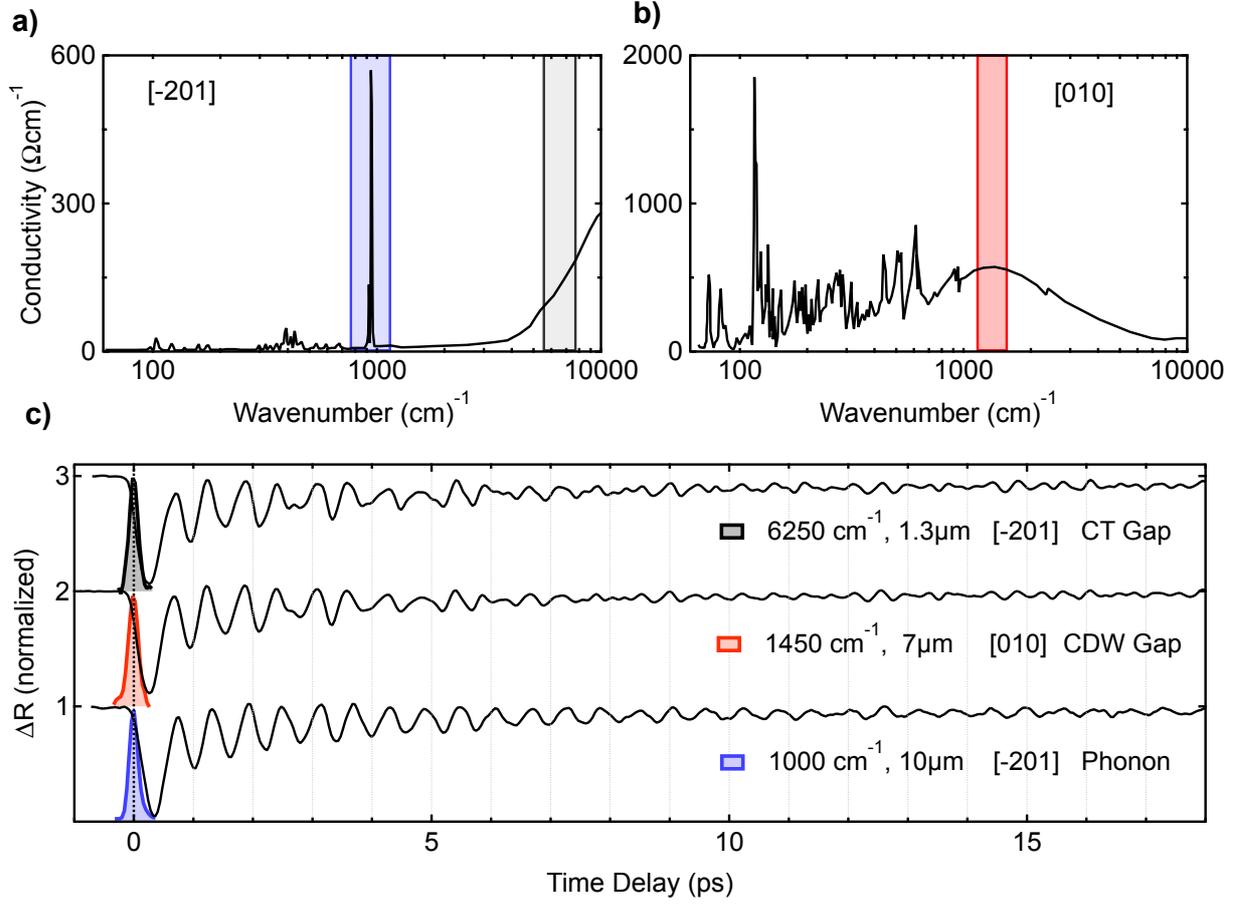

**Fig. 2:** Optical conductivity $\sigma_1$ of $K_{0.3}MoO_3$ below the transition temperature $T_C$. (a) The material is insulating along [-201] with a 1.3-eV charge-transfer gap, and the Mo-O stretching mode appears as a sharp peak at 1000 cm$^{-1}$. (b) The CDW gap is formed along the [010] chain direction. The shaded regions denote the frequencies to which the optical excitation pulses were tuned: infrared-active phonon mode (blue), electronic excitation across the charge-transfer gap (black), and excitation of quasi-particles above the CDW gap (red). (c) Transient reflectivity changes recorded at 800 nm probe wavelength (normalized) for the three excitation schemes in a regime of weak perturbation. Zero time delay was determined from the simultaneously measured four-wave mixing signal at frequency $2\omega_{probe} + \omega_{pump}$, which yielded the pump pulse envelopes shown in the Figure (shaded).

**Figure 3**

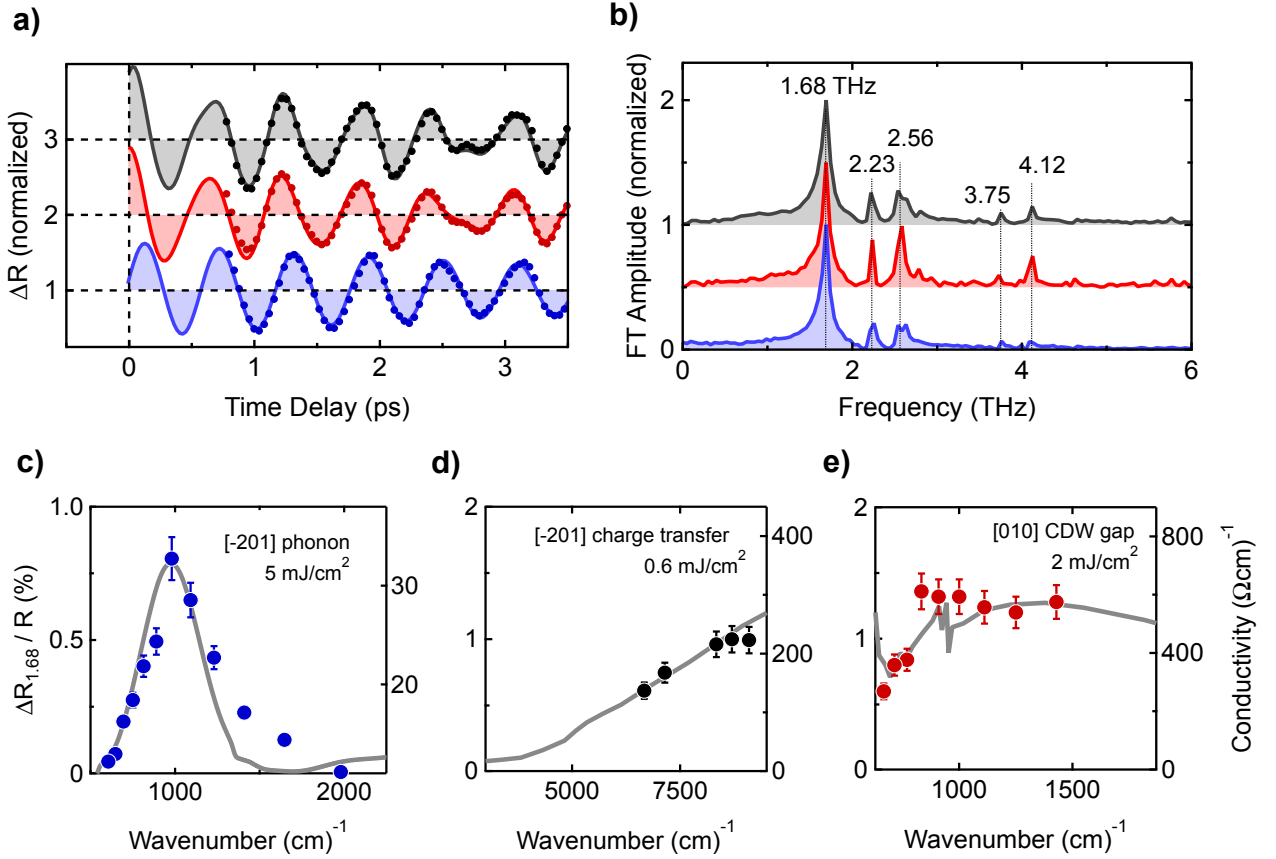

**Fig. 3:** (a) Normalized oscillatory response extracted from the data plotted in Figure 2(c). The different optical excitations are color coded as described in Figs. 2(a) and (b). The extrapolation of the oscillatory response to time zero, fitted with a sum of three exponentially decaying harmonic oscillators, shows a sine-like phase for lattice excitation and a cosine-like phase for the electronic excitations. (b) Fourier transforms of the oscillatory responses, showing the CDW amplitude mode at 1.68 THz and four zone-folded phonon modes at higher frequencies. (c-e) Pump frequency dependent magnitudes of the coherent amplitude mode oscillations, again for the three different excitation schemes. The optical conductivity in the respective frequency ranges and along the respective directions, convolved with the excitation pulse bandwidth is plotted for comparison.

**Figure 4**

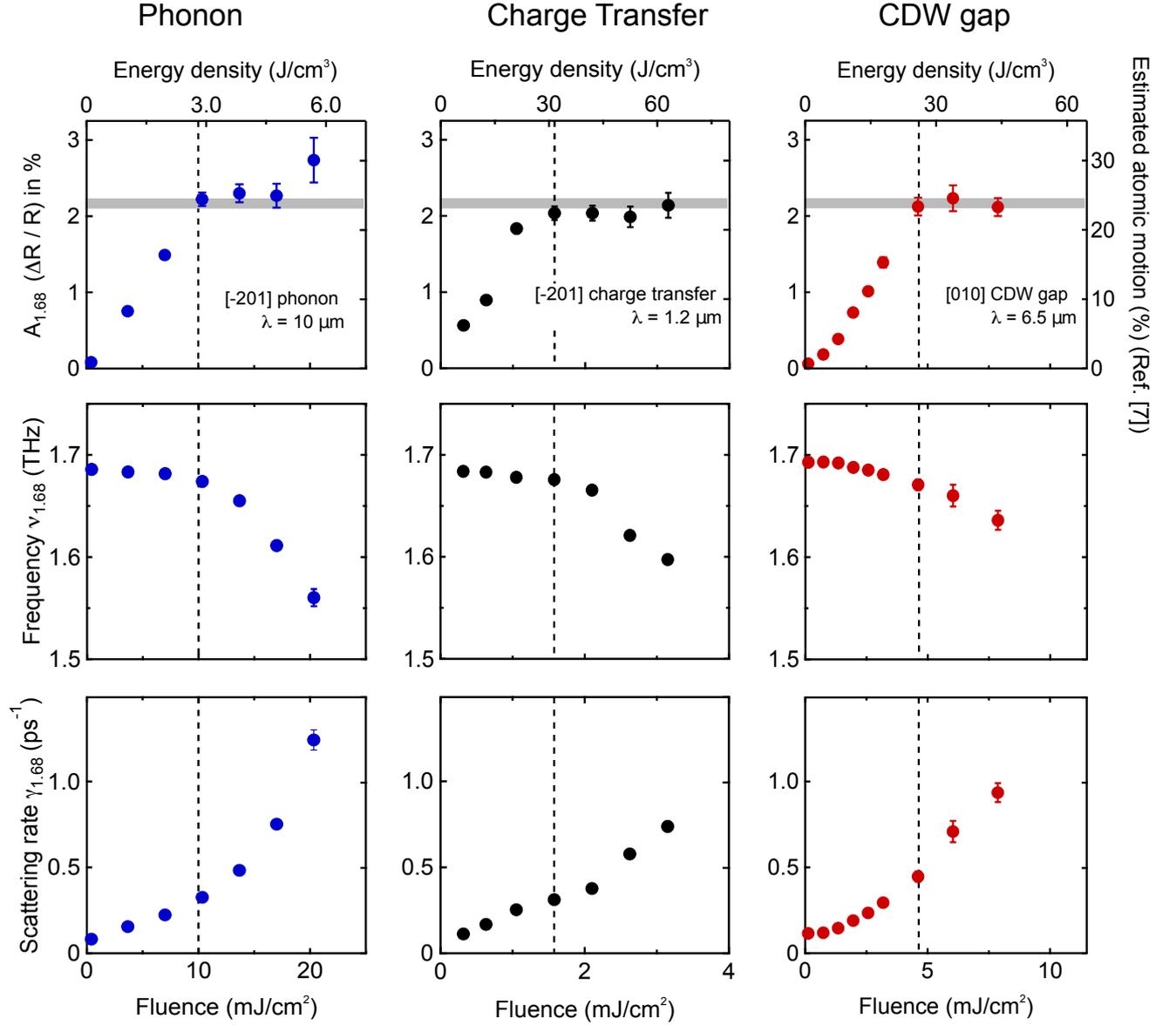

**Fig. 4:** Fluence dependence of the amplitude, frequency and scattering rate of the coherent oscillations of the amplitude mode for resonant phonon excitation (blue) and for electronic excitations above the charge transfer gap (black) as well as across the CDW gap (red). The corresponding energy density is given in the top axis of the upper panels. In all cases, the oscillation amplitude saturated at ~2% $\Delta R/R_0$, which corresponds to atomic displacements of approximately 20% of the lattice distortion induced by the CDW formation at equilibrium as estimated from Refs. 6 and 7 (scale on the right side of upper panel, see text).

**Figure 5**

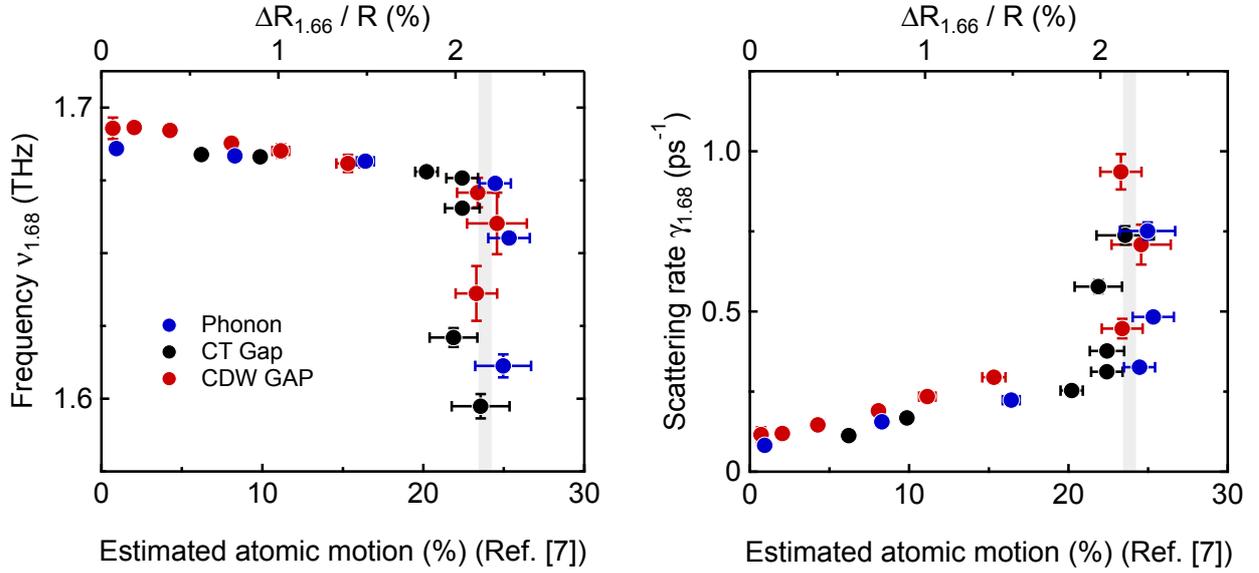

**Fig. 5:** Frequency and scattering rate of the charge density wave amplitude mode, extracted from the data shown in Fig. 4 and plotted against the measured oscillation amplitude. The data fall on a universal curve for the different excitation schemes. The amplitude of the coherent oscillations is given in % ($\Delta R/R_0$) at 800 nm (upper scale) and in the relative change of the lattice distortion, induced by the CDW formation (lower scale).